\documentstyle[12pt,epsf]{article}
\newcommand{\bee}{\begin{eqnarray}}
\newcommand{\eee}{\end{eqnarray}}
\newcommand{\be}{\begin{equation}}
\newcommand{\ee}{\end{equation}}

\begin{document}
\baselineskip = 14pt

\begin{center}{ 
\Large  \bf In-medium spin-orbit NN potential in the Skyrme model}\\[1cm]
A.P.~Kobushkin\\[0.5cm]
{\it Research Center for Nuclear Physics, Osaka University, Osaka 567-0047, Japan}\\
and\\
{\it Bogolyubov Institute for Theoretical Physics, 03143 Kiev, Ukraine}
\end{center}

\vspace{2.cm}
\begin{center}
\large \bf Abstract 
\end{center}
It is considered how spin-orbit potential of two nucleons is changed when one of them is
embedded in nuclear matter. It is emphasized that there appears new type of potential  
antisymmetric under nucleon spins which is proportional to mass difference between 
free and embedded nucleons. The longest range behavior of such 
potential is estimated.\\[2.cm]  

Study of medium effects on nucleon properties is one of hot problems in the modern 
nuclear physics. A direct means to investigate this problem may be realized in
experiments on nucleon quasifree scattering in nuclear  matter at 0.5~---~1 GeV 
energy scale \cite{RCNP}. 

In the present note basing on the Skyrme model we consider trends of modifications
in the NN spin-orbit potential when one of interacting nucleons is embedded in nuclear 
matter (Fig.~1). In this case the nucleon in matter changes its properties (mass, momentum of
inertia, mean squared radius, etc.) and it becomes non identical to the
free nucleon.
 
The Skyrme model provides a compact model for baryons considered as topological 
solitons of meson field. Besides a fact that it takes explicitly into account only meson 
degrees of freedom the model has deep relation with fundamental theory of strong 
interaction, QCD.

The main success of the Skyrme model is connected with calculations of free baryon 
properties \cite{ANW}. It also explains some important features of the NN 
interaction in vacuum \cite{SkyrmeNP}-\cite{RiskaD}.

Properties of a skyrmion in medium were investigated by number of authors
(see \cite{Meissner}-\cite{RahimovMKY} and references therein). Qualitatively many of 
such properties can be obtained from simple scaling rules suggested by Brown and 
Rho \cite{Brown}: in the chiral limit there is only one dimension value, the pion 
decay constant $F_{\pi}$, which is changed in medium as follows
\be
F_{\pi} \to F_{\pi}^{\ast}=F_{\pi}\left( 1-\kappa\frac{\rho}{\rho_0}\right)
\equiv \xi F_{\pi}.
\label{1}
\ee
Here and later  asterisk indicates an observation in medium, $\rho$ denotes nuclear density
and $\rho_0 = 0.16\ {\rm fm}^{-3}$. All dimensional physical quantities are expressed
by appropriate power for the $ F_{\pi}^{\ast}$ and the scaling rules read \cite{Meissner}
\be
\frac{M^{\ast}}{M}= \frac{F_{\pi}^{\ast}}{F_{\pi}},\
\frac{I^{\ast}}{I}=\frac{F_{\pi}}{F_{\pi}^{\ast}},\ {\rm etc.}
\label{2}
\ee
In (\ref{2}) $M$ is a free skyrmion mass with quantum numbers of the nucleon, $I$ is its
momentum of inertia. In turn the pion mass was estimated to be practically independent on 
the nuclear density \cite{BMeissner}.

For a skyrmion in nuclear matter the profile function is changed $f(R) \to f^{\ast}(R,\xi)$.
For simplicity we shall use a phenomenologic $\xi$-dependence of the profile function 
basing on the following arguments. One may assume that at distance between the skyrmion 
centers $R\ll m_{\pi}^{-1}$ the chiral limit is fulfilled and the modified profile function 
is determined from scaling low    
\be
f(R)=\Theta(e_S F_{\pi}R) \to f^{\ast}(R)=\Theta(e_S F^{\ast}_{\pi}R),\
R\ll m_{\pi}^{-1},
\label{3}
\ee
where $e_S$  is the Skyrme constant. But at region  $R\gg m_{\pi}^{-1}$ it must obey  
the Yukawa low and
\be
f(R)= f^{\ast}(R) \sim \frac1r \exp(-m_{\pi} R),\
R\gg m_{\pi}^{-1}.
\label{14}
\ee
Thus one may accept the following $\xi$-dependence of the profile function, which reproduces 
this two asymptotic regimes
\be
f^{\ast}(R)= f(\kappa(\xi,R)R),\ \kappa =\frac{R +\xi a m^{-1}_\pi}{R+ a m^{-1}_\pi}, 
\label{15}
\ee 
where $a$ is an adjustable parameter which takes into account difference between
nuclei size and size of the nucleon.

Due to the scaling low the problem seems to be similar to that of potential for
two independently breathing skyrmions \cite{KalbE},\cite{Kob}. Between different
important effects of the breathing mode one has to mention that it generate new
type of the spin-orbit potential, which is proportional
to $(\vec \sigma^{(1)}- \vec \sigma^{(2)})\vec l$, where the  $(\vec \sigma^{(1)}$
and  $\vec \sigma^{(2)}$ are spin Pauli matrices of the skyrmions) \cite{Kob}. From identity
of the free nucleons it is obvious that this potential does not contribute  to the 
$NN \to NN$ transition in vacuum and is realized as spin-orbit transition potential in 
the $NN \to NN(1440)$ and so on channels. Here we will discuss how the
antisymmetric spin-orbit structure is realized in the NN potential when one of the 
nucleons is  embedded in nuclear matter. 

We start from the product approximation \cite{SkyrmeNP}
\be
U(\vec r,t;\vec R^{(1)},\vec R^{(2)})=
U_1(\vec r - \vec R^{(1)};t)U_2(\vec r - \vec R^{(2)};t),
\label{4}
\ee
which is justified when the separation, $R=|\vec R^{(1)}-\vec R^{(2)}|$, between the center of the 
two skyrmions,  
$\vec R^{(1)}$ and $\vec R^{(2)}$, exceeds the skyrmion size. The other notations 
are as follows: $(\vec r,t)$ is the space-time coordinate of the soliton field and
$U_1(\vec r^{\,(1)} ;t)$  and 
$U_2(\vec r^{\,(2)} ;t)$ are the adiabatically rotated ``hedgehogs''
\bee
&&U_i(\vec r^{\,(i)};t) = A_i(t) U^0_i(\vec r^{\,(i)}) A^{\dag}_i(t),
\label{5}\\
&&U^0_1(\vec r^{\,(1)}) = \exp[i\vec \tau \hat r^{(1)} f(r^{(1)})],\
U^0_2(\vec r^{\,(2)})  =  \exp[i\vec \tau \hat r^{(2)} f^{\ast}(r^{(2)})],
\label{6}
\eee
where $\vec r^{\,(i)} \equiv \vec r-\vec R^{\,(i)}$ and $A_i(t)$ are $2 \times 2$
rotation matrices. Adapting arguments of \cite{Kob} 
one gets the following structure of isospin dependent spin-orbit potential
\be
V_{T=1,ls}(R,\xi) = 
(\vec \tau^{\,(1)}\vec \tau^{\,(2)})
\left[ (\vec \sigma^{(1)} + \vec \sigma^{(2)})\vec l V^{(+)}_{T=1}(R,\xi)+
(\vec \sigma^{(1)} - \vec \sigma^{(2)})\vec l V^{(-)}_{T=1}(R,\xi)
\right].
\label{7}
\ee 
The longest range behavior of the potentials $V^{(\pm)}_{T=1}(R,\xi)$ comes from 
the quadratic part of the Skyrme lagrangian 
\be
V^{(\pm)}_{T=1}(R,\xi) = -\frac{F_{\pi}F^{\ast}_{\pi}}{72 R^2}
\int d^3r
\left[
\frac{(\vec r^{\,(1)}\cdot \vec R)}{r^{\,(1)\,2} I^{\ast }M_N} \pm
\frac{(\vec r^{\,(2)}\cdot \vec R)}{r^{\,(2)\,2} I M_N^{\ast}}
\right]
\sin^2 f(r^{\,(1)})\sin^2 f^{\ast}(r^{\,(2)}),
\label{8}  
\ee
where for symmetry we replace $F^2_{\pi} \to F_{\pi}F^{\ast}_{\pi} $. If both skyrmions are 
in vacuum the antisymmetric potential $V^{(-)}$ vanishes. Note that antisymmetric 
spin-orbit potential also appears for the hyperon-nucleon interaction  which is
proportional to mass difference between hyperon and nucleon \cite{deSwart}. 

When the distance between skyrmions is larger than the skyrmion size the integral
in (\ref{8}) can be divided to two parts in vicinity of each skyrmion 
\be
\int d^3 r \approx \int d^3r^{\,(1)} +  \int d^3r^{\,(2)}. 
\label{10}
\ee
For the first integral one may replace $\vec r^{\,(2)} \to \vec R$ and for the
second integral  one may replace $\vec r^{\,(1)} \to \vec R$
(see \cite{Nyman}) and the expression (\ref{8}) becomes
\bee
&&V^{(\pm)}_{T=1}(R,\xi) =
 -\frac{F_{\pi}F^{\ast}_{\pi}}{72 R^2}
\left\{
\pm \frac{\sin^2 f^{\ast}(R)}{I^{\ast} M}\int d^3 r^{\,(1)} \sin^2
f(r^{\,(1)}) +
\right. \nonumber \\
&+& \left. 
    \frac{\sin^2 f(R)}       {I M^{\ast}}\int d^3 r^{\,(2)} \sin^2 f^{\ast}(r^{\,(2)})
\right\}.
\label{11}
\eee
In (\ref{11}) one may put approximately
\be
\int d^3 r^{\,(1)}  \sin^2 f(r^{\,(1)}) \approx \frac{6}{F^2_{\pi}}I\
{\rm and}\ 
\int d^3 r^{\,(2)}  \sin^2 f^{\ast}(r^{\,(2)}) \approx \frac{6}{F^{\ast\, 2}_{\pi}}I^{\ast}
\label{12}
\ee
and in closed form the $V^{(\pm)}_{T=1}(R)$ potentials read
\be
V^{(\pm)}_{T=1}(R) \approx
-\frac{\xi}{12 R^2 M}\left[
\xi^{-4}\sin^2 f(R) \pm \xi^4\sin^2 f^{\ast}(R)
\right].
\label{13}
\ee
Eq.~(\ref{13}) gives very simple connection between the isospin-dependent spin-orbit
potential for the nucleons in vacuum, $V^{\rm vac}_{T=1,ls}(R)$,  and the potentials 
$V^{(\pm)}_{T=1}(R)$
\be
V^{(\pm)}_{T=1}(R)=
\frac 12 \left[
\xi^{-3}V^{\rm vac}_{T=1,ls}(R) \pm \xi^5 V^{\rm vac}_{T=1,ls} (\kappa(\xi,R)\cdot R)
\right].
\label{20}
\ee
Numerical calculations show that at not very high density $\rho$ one can neglect in 
(\ref{20}) by the $\xi$-dependence of the potential, 
$V^{\rm vac}_{T=1,ls}(\kappa(\xi,R)\cdot R) \approx V^{\rm vac}_{T=1,ls}(R)$, and 
the antisymmetric potential comes only from differences between mass and momentum of 
inertia  of the nucleon in medium and in vacuum
\be
V^{(\pm)}_{T=1}(R) \approx \kappa\frac{\rho}{\rho_0}V^{\rm vac}_{T=1,ls}(R).
\label{20.a}
\ee
In  Fig.~2 we display the potentials $V^{(\pm)}_{T=1}(R)$ obtained from Eq.~(\ref{20})
for the Paris potential \cite{Paris} with parameters $a=2$ and $\xi=0.7$. We also 
compare them with the isospin independent spin-orbital Paris potential 
$V_{T=1,ls}^{\rm vac}(R)$ in vacuum.

Of course an antisymmetric spin-orbit potential appears also for the isospin-independent
spin-orbit potential. The main reason is the same as for the isospin-dependent
potential. Unfortunately even for the free nucleons the Skyrme model gives results
for this potential which disagrees qualitatively with results from realistic 
phenomenological models \cite{RiskaD}. This is related to the fact that the 
isospin-independent spin-orbit potential comes from the quartic term in the Skyrme 
lagrangian which fails to reproduce attractive interaction at low and intermediate 
distances \cite{RiskaD}. Thus direct calculation of the antisymmetric 
spin-orbit potential in the Skyrme model seems to be senseless and
we estimate only tendency of this potential from arguments similar to that were used
for (\ref{20.a})
\be
V_{T=0}^{(-)}\approx \kappa\frac{\rho}{\rho_0} V^{\rm vac}_{T=0,ls}(R),
\label{21}
\ee
where $ V^{\rm vac}_{T=0,ls}(R)$ is isospin-independent spin-orbit NN potential in vacuum. 

In summary, the Skyrme model predicts antisymmetric spin-orbit structure for the NN
potential, when one of interacting nucleons is embedded in nuclear matter. The main 
reason for this potential is connected with difference between free and effective mass 
and momentum of inertia of nucleon. Possible experimental observation of effects 
connected with this potential may give us a direct answer on a question how strongly
nuclear matter changes properties of the nucleon.
\\[0.5cm]
The author would like to thank Prof.~T.~Noro for useful discussions which strongly
stimulated this work.

\newpage
\begin{center}
{\large \bf
Figure Captions
}
\end{center}
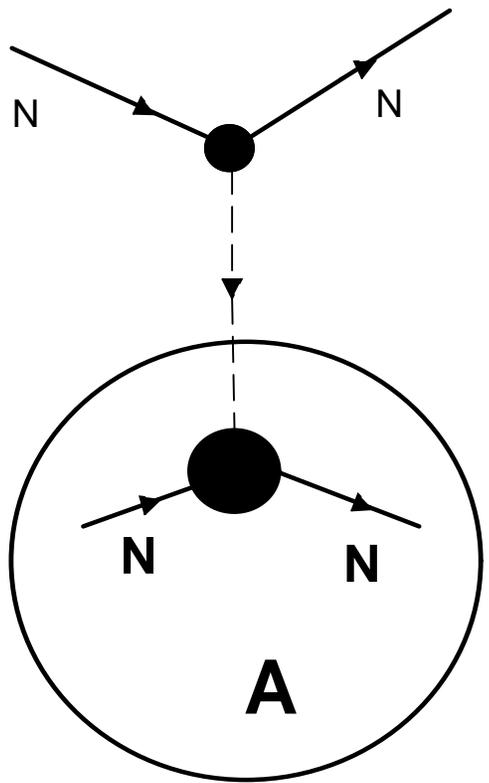
Figure 1. Nucleon embedded in nuclear  matter changes mass,
size and momentum of inertia.
\\[1.cm]
Figure 2. Modified spin-orbit isospin-dependent potential: the 
antisymmetric potential $V^{(-)}_{T=1}$ is shown by full line,
the symmetric potential  $V^{(+)}_{T=1}$ --- dashed line and the
potential in vacuum $V^{\rm vac}_{T=1,ls}$ --- dashed-dotted line.


\newpage
\begin{thebibliography}{99}
\bibitem{RCNP}
K.~Hatanaka {\it et al.}, Phys. Rev. Lett. {\bf 78} (1997) 1014;
T.~Noro {\it et al.}, Nucl. Phys. {\bf A663\&664} (2000), 517c;
T.~Noro {\it et al.}, in Proc. of the RCNP Inter. Symp. on Nuclear Responses and
Medium Effects, Osaka, Japan, edited by T.~Noro et al., Universal Academy 
Press, Inc. Tokyo, Japan, p.167.
\bibitem{ANW}
G.S.~Adkins, C.R.~Nappi and E.~Witten, Nucl. Phys. {\bf B228} (1983) 552;
G.S.~Adkins and C.R.~Nappi, {\it ibid}.{\bf B233} (1984) 109;
I.~Zahed and G.~Brown, Phys. Rep. {\bf 142} (1986) 1.
\bibitem{SkyrmeNP}
T.H.R.~Skyrme, Nucl.Phys. {\bf 31} (1962) 556.
\bibitem{JJP}
A.~Jackson, A.D.~Jackson and V.~Pasquir, Nucl. Phys. {\bf A432} (1985) 567.
\bibitem{NRiska1}
E.M.~Nyman and D.O.~Riska, Phys. Lett.{\bf B175} (1986) 392.
\bibitem{NRiska2} 
E.M.~Nyman and D.O.~Riska, Physica Scripta {\bf 34} (1986) 533.
\bibitem{RiskaN1}
D.O.~Riska and E.M.~Nyman, Phys. Lett.{\bf B183} (1987) 7.
\bibitem{RiskaD}
D.O.~Riska and K.~Dannbom, Physica Scripta {\bf 37} (1988) 7.
\bibitem{Meissner}
U.-G.~Meissner, Nucl. Phys. {\bf A503} (1989) 801.
\bibitem{KalbFE}
G.~Kalbermann, L.L.~Frankfurt and J.E.~Eisenberg, Phys. Lett. {\bf B329} (1994) 164.
\bibitem{Kalb}
G.~Kalbermann, Nucl.  Phys. Lett. {\bf A612} (1997) 359.
\bibitem{RahimovOMK}
A.~Rakhimov {\it et al.}, Phys. Lett. {\bf B378} (1996) 12.
\bibitem{RahimovMKY}
A.~Rakhimov {\it et al.}, Phys. Rev. {\bf C58} (1998) 1738.
\bibitem{Brown}
G.E.~Brown and M.~Rho, Phys. Rev. Lett. {\bf 66} (1991) 2720;
G.~Brown, Prog. Theor. Phys. Suppl. {\bf 91} (1987) 85;
Nucl. Phys. {\bf A488} (1988) 689~c;
U.-G.~Meissner and V.~Benard, Comm. Nucl. Part. Phys. {\bf 19} (1989) 67.
\bibitem{BMeissner}
V.~Bernard and U.-G.~Meissner, Nucl. Phys. {\bf A489} (1988) 647.
\bibitem{KalbE}
G.~Kalbermann and J.M.~Eisenberg, Nucl. Phys. {\bf A500} (1989) 589.
\bibitem{Kob}
A.P.~Kobushkin, Physica Scripta {\bf 55} (1997) 27.
\bibitem{deSwart}
M.M.~Nagels, T.A.~Rijken, and J.J.~de~Swart, Phys. Rev. {\bf D15} (1977) 2547.
\bibitem{Nyman}
E.M.~Nyman, Phys. Lett. {\bf B142} (1987) 388.
\bibitem{Paris}
M.~Lacombe {\it et al.}, Phys. Rev. {\bf C21} (1980) 861.
\end{thebibliography}
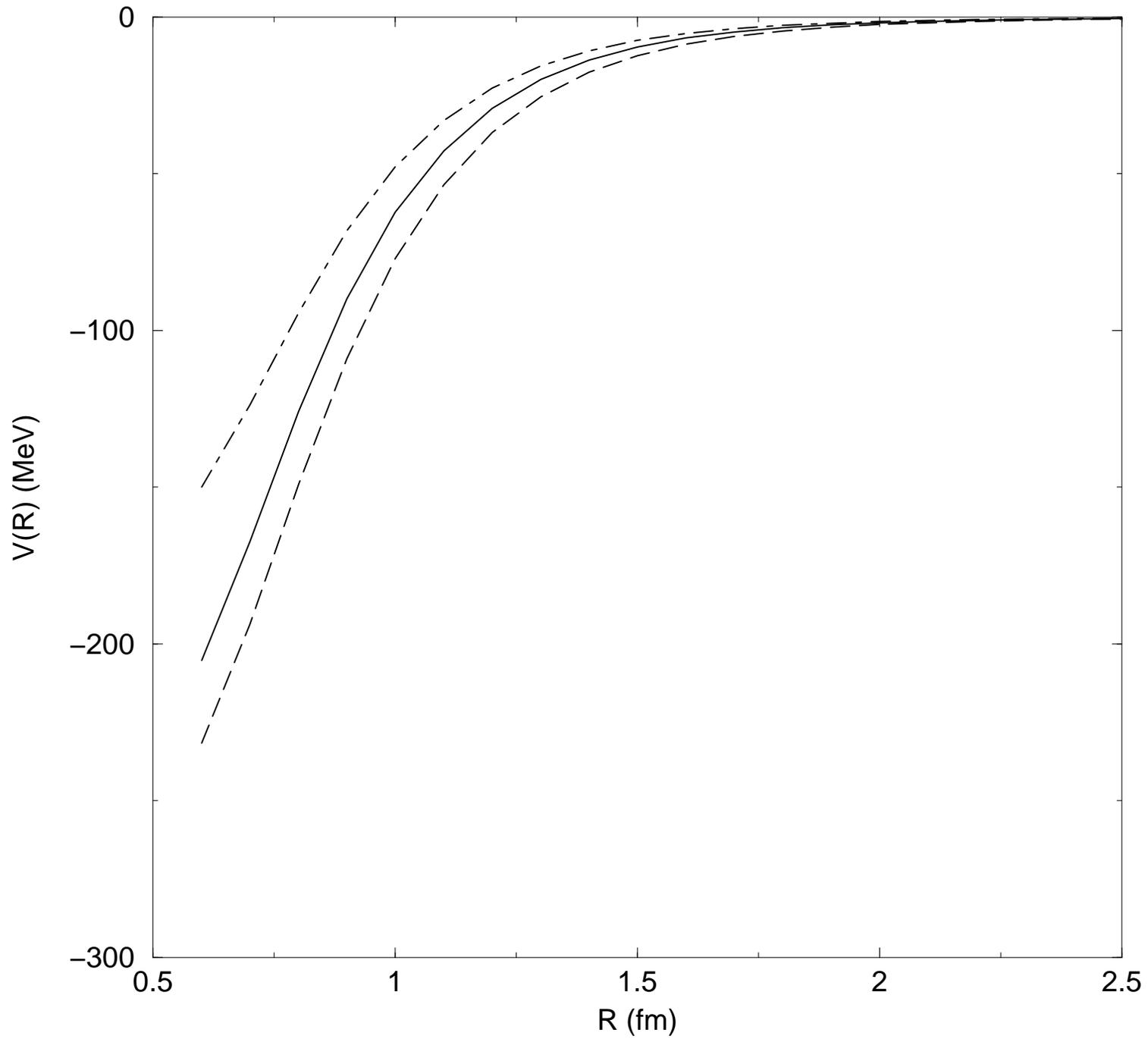
\end{document}